\documentclass[final,5p,numbers,sort&compress,twocolumn]{elsarticle}

\usepackage{amssymb}
\usepackage{xcolor}
\usepackage{graphicx}
\usepackage{dcolumn}
\usepackage{bm}
\usepackage{longtable}
\usepackage{float}
\usepackage{threeparttable}
\usepackage[colorlinks = true, citecolor=blue,urlcolor=blue,linkcolor=blue]{hyperref}
\usepackage{color} 

\usepackage{booktabs}
\usepackage{supertabular}%

\usepackage{url}
\usepackage{array}
\usepackage{dblfloatfix}
\usepackage{tabu}
\usepackage{amsmath}
\usepackage{mathtools, cuted}
\usepackage[english]{babel}
\usepackage[output-decimal-marker={.},exponent-product=\cdot]{siunitx}

\journal{Physics Letters B}

\definecolor{mbscolor}{rgb}{0.60, 0.0, 0.65}

\begin{document}

\begin{frontmatter}

\title{Precision mass measurements in the zirconium region\\ pin down the mass surface across the neutron midshell at $N=66$}

\author[jyv,bor]{M.~Hukkanen\corref{cor}}
\ead{marjut.h.hukkanen@jyu.fi}
\cortext[cor]{Corresponding authors}
\author[ulb]{W.~Ryssens}
\author[bor]{P.~Ascher}
\author[lyon]{M.~Bender}
\author[jyv]{T.~Eronen}
\author[bor]{S.~Gr\'evy}
\author[jyv]{A. Kankainen\corref{cor}}
\ead{anu.kankainen@jyu.fi}
\author[jyv]{M.~Stryjczyk}
\author[jyv]{O.~Beliuskina}
\author[jyv]{Z.~Ge}
\author[kul]{S.~Geldhof\fnref{ganil}}
\author[bor]{M.~Gerbaux}
\author[jyv]{W.~Gins}
\author[bor]{A.~Husson}
\author[jyv]{D.A.~Nesterenko}
\author[jyv]{A.~Raggio}
\author[jyv]{M.~Reponen}
\author[jyv]{S.~Rinta-Antila}
\author[jyv,liverpool]{J.~Romero}
\author[jyv]{A.~de Roubin\fnref{lpc}}
\author[jyv]{V.~Virtanen}
\author[jyv]{A.~Zadvornaya\fnref{edinburgh}}

\address[jyv]{University of Jyvaskyla, Department of Physics, Accelerator laboratory, P.O. Box 35(YFL) FI-40014 University of Jyvaskyla, Finland}
\address[bor]{Universit\'e de Bordeaux, CNRS/IN2P3, LP2I Bordeaux, UMR 5797, F-33170 Gradignan, France}
\address[ulb]{Institut d'Astronomie et d'Astrophysique, Universit\'e Libre de Bruxelles, Campus de la Plaine CP 226, 1050 Brussels, Belgium}
\address[lyon]{Universit\'e Claude Bernard Lyon 1, CNRS/IN2P3, IP2I Lyon, UMR 5822, F-69622 Villeurbanne, France}
\address[kul]{KU Leuven, Instituut voor Kern- en Stralingsfysica, B-3001 Leuven, Belgium}
\address[liverpool]{Department of Physics, Oliver Lodge Laboratory, University of Liverpool, Liverpool, L69 7ZE, United Kingdom}

\fntext[ganil]{Present address:  GANIL, CEA/DRF-CNRS/IN2P3, B.P. 55027, 14076 Caen, France}
\fntext[lpc]{Present address: LPC Caen, Normandie Univ., 14000 Caen, France}
\fntext[edinburgh]{Present address: University of Edinburgh, Edinburgh EH9 3FD, United Kingdom}

\begin{abstract}
Precision mass measurements of $^{104}$Y, $^{106}$Zr, $^{104,104m,109}$Nb, and $^{111,112}$Mo have been performed with the JYFLTRAP double Penning trap mass spectrometer at the Ion Guide Isotope Separator On-Line facility. The order of the long-lived states in $^{104}$Nb was unambiguously established. The trend in two-neutron separation energies around the $N=66$ neutron midshell appeared to be steeper with respect to the Atomic Mass Evaluation 2020 extrapolations for the $_{39}$Y and $_{40}$Zr isotopic chains and less steep for the $_{41}$Nb chain, indicating a possible gap opening around $Z=40$.  
The experimental results were compared to the BSkG2 model calculations performed with and without vibrational and rotational corrections. All of them predict two low-lying minima for $^{106}$Zr. 
While the unaltered BSkG2 model fails to predict the trend in two-neutron separation energies, selecting the more deformed minima in calculations and removing the vibrational correction, the calculations are more in line with experimental data. The same is also true for the $2^+_1$ excitation energies and differences in charge radii in the Zr isotopes. The results stress the importance of improved treatment of collective corrections in large-scale models and further development of beyond-mean-field techniques. 
\end{abstract}

\end{frontmatter}


\section{\label{sec:intro}Introduction}

The region of neutron-rich refractory nuclei around ${A\approx100}$ is known for shape coexistence and rapid and abrupt changes in ground-state deformation. Nuclear spectroscopy studies have indicated coexisting spherical, oblate and prolate shapes in the region (see e.g. Refs.~\cite{Heyde2011,Garrett2022}) while laser spectroscopy studies have revealed a sudden increase in the mean-square charge radii at ${N\approx60}$ for the isotopic chains from Rb (${Z=37}$) to Nb (${Z=41}$) \cite{Thibault1981,Procter2015,Silverans1988,Cheal2007,Campbell2002,Cheal2009}. A sharp drop in the excitation energies of the first ${2^+}$ states at ${N=60}$ is an observational signal of a change in deformation. The drop is most pronounced in the Sr (${Z=38}$) and Zr (${Z=40}$) chains while being somewhat smoother for the Kr ($Z=36$), Mo (${Z=42}$) and Ru (${Z=44}$) chains \cite{Garrett2022}. Changes in nuclear structure are also reflected in nuclear binding energies with two-neutron separation energies ($S_{2n}$) showing a kink at ${N\approx60}$ from Rb to Mo \cite{AME20}. This behavior correlates with the observed onset of deformation, see e.g. Fig. 15 in Ref.~\cite{Kankainen2012}.

While the ${N\approx60}$ region has been extensively studied for the refractory elements with ${Z\approx40}$, not much is known about the structure of these elements around the ${N\approx66}$ midshell. In other neutron midshell regions interesting phenomena have been observed with the most striking one present in the neutron-deficient Hg isotopes (${Z=80}$) where the mean-square charge radii exhibit a unique odd-even staggering at $N\approx104$ \cite{Kuhl1977,Marsh2018}. Other examples include a very large deformation observed for ${Z\approx N \approx 40}$ \cite{Llewellyn2020}, as well as the Cd isotopes (${Z=48}$) where the low-lying collective states reach their minimum at $N\approx66$ \cite{Juutinen1996,Garrett2022}. While these structural changes have been clearly observed via spectroscopy experiments, they do not necessarily introduce observable effects in the ground-state binding energies.

Laser spectroscopy has not yet reached exotic Zr and Mo isotopes at ${N\approx66}$. However, $\gamma$-ray spectroscopy experiments at the Radioactive Isotope Beam Factory have reported that the excitation energies of the ${2^+}$ states for the Zr isotopes reach a minimum at ${N=64}$ and start to very slowly increase again for ${N>64}$ \cite{Sumikama2011,Paul2017}, similarly to the Mo isotopic chain \cite{ENSDF}. This suggests that the maximum quadrupolar deformation is reached at $N=64$. It is worthwhile to note that also $^{110}$Zr at ${N=70}$ is still strongly deformed, with ${E(2_1^+)=185(11)}$~keV and the excitation energy ratio ${R=E(4_1^+)/E(2_1^+)=3.1(2)}$ \cite{Paul2017}. 

Despite the efforts, the mass surface in the ${N\approx66}$ midshell region has remained poorly known, in particular for the Y (${Z=39}$) and Zr isotopic chains. Although the masses of $^{105}$Y and $^{106}$Zr were reported in Ref.~\cite{Knobel2016}, they were deemed anomalous \cite{Huang2021}. Consequently, they were removed from the Atomic Mass Evaluation 2020 (AME20) and replaced by extrapolated values.
In the lighter isotopic chains of Rb and Sr, the deformation has been found to extend toward larger neutron numbers \cite{deRoubin2017}. No subshell gap or indications of change in deformation in the ${N=66}$ region have been found in these isotopic chains via mass measurements \cite{Mukul2021}. In the heavier isotopic chains, our recent mass measurements show that there are no drastic changes in the $S_{2n}$ values of the Ru (${Z=44}$) \cite{Hukkanen2023b} and the Rh (${Z=45}$) \cite{Hukkanen2023} isotopic chains up to $N{\approx70}$. However, the first Brussels-Skyrme-on-a-Grid (BSkG1) \cite{Scamps2021} mass model calculations predicts that the neutron-rich Ru and Rh isotopes at ${N\approx66}$ are triaxially deformed and a structural change could affect the $S_{2n}$ trends at higher neutron numbers \cite{Hukkanen2023,Hukkanen2023b}.

For the Zr ground states, axial mean-field approaches have mainly predicted axially-deformed shapes in the ${N\approx60-72}$ region, see e.g. Ref.~\cite{chimanski2023}. Not so many models have taken into account triaxial degrees of freedom. However, they were included in the 5-Dimensional Collective Hamiltonian (5DCH) \cite{Delaroche2010} with the Gogny D1S effective interaction \cite{Decharge1980,Berger1991}, as well as in the projected configuration mixing model (PCM) \cite{Bender2008,Borrajo2015} that uses either the D1S or SLyMR0 force \cite{Bally2014,Sadoudi2013}. Within these models, the ground and the $2_{1,2}^+$ states in $^{110}$Zr are triaxial and they are built on rather $\gamma$-soft potential energy surfaces, see the supplemental information in Ref.~\cite{Paul2017}. These predictions are also supported by the large-scale Monte Carlo shell-model calculations which have predicted triaxial excited bands for Zr isotopes with ${N\geq66}$ \cite{Togashi2016}. The strong deformation continuing up to $^{110}$Zr has also been experimentally verified \cite{Paul2017}. 

In this Letter, we report on Penning-trap mass measurements in the ${N\approx66}$ region and present the first high-precision mass measurement of $^{104}$Y (${Z=39}$, ${N=65}$) and $^{106}$Zr (${Z=40}$, ${N=66}$). The theoretical interpretation of the results is performed in the framework of the BSkG2 model \cite{Ryssens2022}. This new mass model is based on Skyrme energy-density functional similarly to its predecessor BSkG1 \cite{Scamps2021}. The main improvements in BSkG2 compared to BSkG1 are related to the treatment of time-reversal invariance and incorporation of fission barriers of actinide nuclei \cite{Ryssens2022,Ryssens2023}. Both models allow for triaxial degrees of freedom relevant for the studied region.

\section{\label{sec:exp}Experimental method}

Precision mass measurements were performed at the Ion Guide Isotope Separator On-Line (IGISOL) facility \cite{Moore2013} utilizing the JYFLTRAP double Penning trap \cite{Eronen2012}. The measured isotopes were produced in proton-induced fission by impinging 25-MeV proton beam from the K130 cyclotron on a 15 mg/cm$^2$ $^{\rm nat}$U target. The fission products were stopped in a helium gas cell working at around 300~mbar and extracted using a sextupole ion guide \cite{Karvonen2008}. The secondary beam was accelerated to 30$q$ keV and mass separated based on their mass-to-charge ratio by a 55 degree dipole magnet. Then, the ions were transported to the buffer gas-filled radio-frequency quadrupole \cite{Nieminen2001}, cooled, bunched and sent to JYFLTRAP. 

In the first trap, the ions were cooled, centered and prepared utilizing the mass-selective buffer gas cooling technique \cite{Savard1991}. The selected ions were sent into the second trap where their cyclotron frequency ${\nu_c = qB/(2 \pi m)}$ was measured. The $q/m$ value is the charge-to-mass ratio of the ion and $B$ is the magnetic field strength, which was determined using the reference ions of $^{133}$Cs delivered from the offline ion source \cite{Vilen2020}.

\begin{figure}[t]
\centering
\includegraphics[width=\columnwidth]{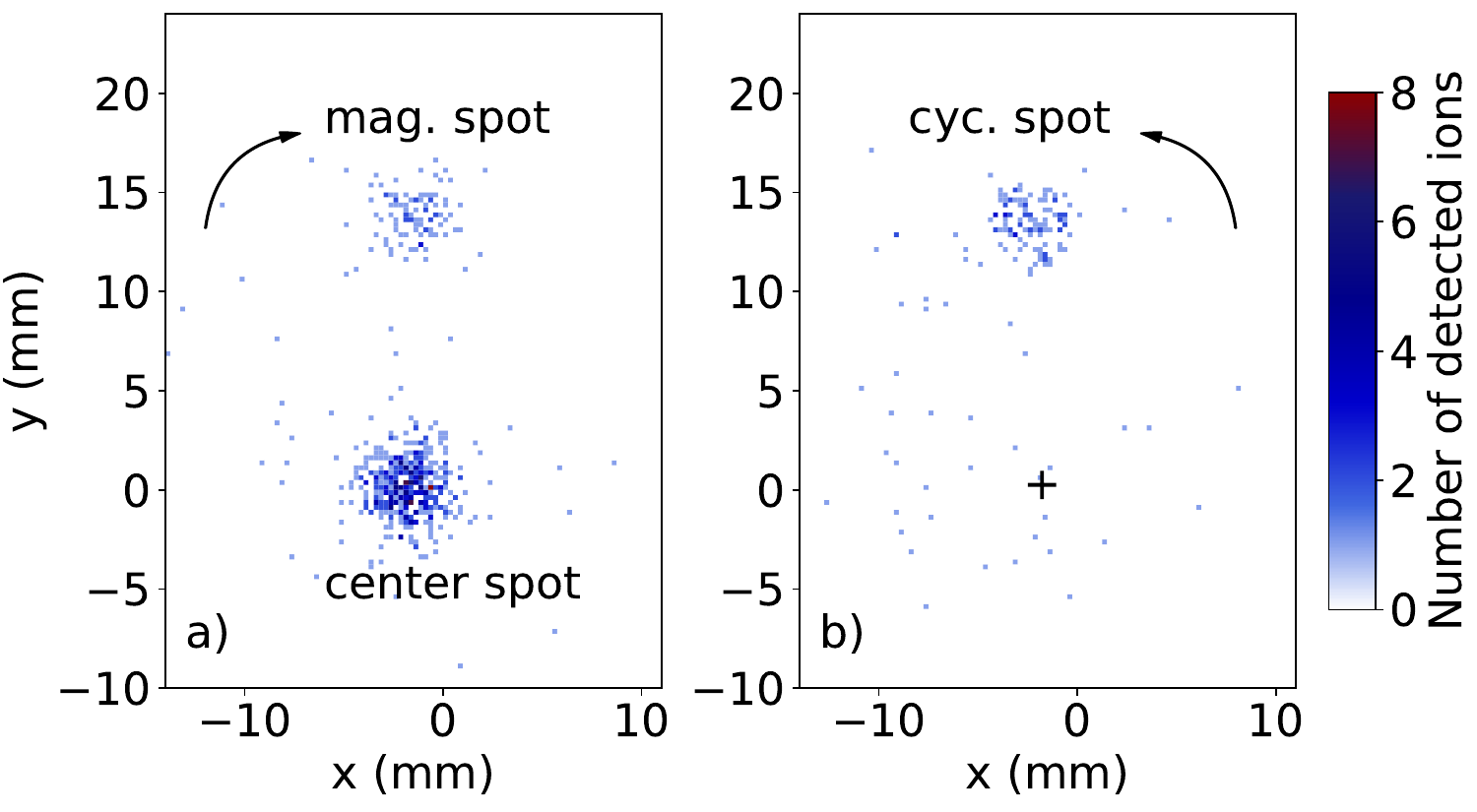}
\caption{\label{fig:106ZrPI-ICR} Ion projection of a) the magnetron, the center and b) the cyclotron phase spots of $^{106}$Zr$^+$ ions on the 2D MCP detector collected with $t_{\rm acc} = 54$~ms. The $+$ symbol in panel b indicates the position of the center spot.}
\end{figure}

The time-of-flight ion-cyclotron-resonance (TOF-ICR) technique \cite{Graff1980, Konig1995} with a $50$ ms quadrupolar excitation was used for $^{109}$Nb and $^{112}$Mo ions while the phase-imaging ion-cyclotron-resonance (PI-ICR) technique \cite{Eliseev2014, Nesterenko2018} was utilized for the other cases. In the TOF-ICR technique, the ions for which the excitation frequency $\nu_{RF}$ matches with their cyclotron frequency $\nu_c$ will have the shortest time of flight from the trap to a micro-channel plate (MCP) detector. In the PI-ICR technique, $\nu_c$ is determined from the angular difference between the projections of the magnetron and cyclotron radial in-trap motion images, measured with respect to the center spot on a position sensitive MCP detector (2D MCP) after a phase accumulation time $t_{\rm acc}$, see Fig. \ref{fig:106ZrPI-ICR}. 

The atomic mass of the ion of interest $M$ was deduced using the frequency ratio $r = \nu_{c,{\rm ref}}/\nu_{c}$ of the measured cyclotron frequencies:
\begin{equation}
M = r\frac{z}{z_{\rm ref}}(M_{\rm ref} - z_{\rm ref} m_e) + z m_e\mathrm{,}
\end{equation}
where $m_e$ is the mass of an electron, $M_{\rm ref}$ is the atomic mass of the reference ion while $z$ and $z_{\rm ref}$ are charge states of the ion of interest and the reference ion, respectively. The measurements were performed with singly-charged ions, except for the isomeric state of $^{104}$Nb for which $z=2$.

Ion-ion interactions were taken into account by either limiting the number of detected ions per bunch or, in the case of $^{104}$Nb, by using a count-rate class analysis \cite{Kellerbauer2003,Nesterenko2021}. To reduce systematic effects due to magnetic field fluctuations, the ion of interest and the reference ion were measured alternately and an extra uncertainty of ${\delta B/B = 2.01(25) \times 10^{-12}\mathrm{~min}^{-1} \times \delta t}$, with $\delta t$ being the time between the measurements, was added. The mass-dependent uncertainty of ${\delta r/r = -2.35(81) \times 10^{-10} / \textnormal{u} \times (M_{\rm ref} - M)}$ and a residual systematic uncertainty of $\delta r/r= 9\times 10^{-9}$ were also taken into account in the analysis. For the PI-ICR measurements the systematic angle error was also accounted for. A detailed description on the systematic uncertainties and measurement patterns at JYFLTRAP can be found in Refs.~\cite{Nesterenko2018,Nesterenko2021}.

\section{\label{sec:results}Results}

\begin{table*}
\centering
\begin{threeparttable}
\caption{\label{tab:results} Frequency ratios $r = \nu_{c,{\rm ref}}/\nu_{c}$ and corresponding mass-excess values (ME) determined in this work using $^{133}$Cs$^+$ ions as a reference. Mass-excess values from AME20 (ME$_{\rm lit.}$) \cite{AME20} and the differences $\mathrm{Diff.}= \mathrm{ME}-\mathrm{ME}_{\rm lit.}$ are added for comparison. $t_{\rm trap}$ is the phase accumulation time for the PI-ICR measurements (excitation time for the TOF-ICR measurements, labeled with $^t$). The half-lives $T_{1/2}$ and spin-parity assignments $I^\pi$ are taken from NUBASE20 \cite{NUBASE20}. The order of the states in $^{104}$Nb is established based on the measurements from this work, see text for details. The state marked with $^x$ is not unambiguously identified, see text for details. \# denotes extrapolated values.}
\begin{tabular}{llllllll}
\hline
Nuclide & $I^{\pi}$ & $T_{1/2}$ & $t_{\rm trap}$ (ms) & $r = \nu_{c,{\rm ref}}/\nu_{c}$ & ME (keV) & ME$_{\rm lit.}$ (keV) & Diff. (keV)\\\hline
$^{104}$Y & ($0^+,1^+)\#$ & 197(4) ms  & 30 & \num{0.78207412(13)} & \num{-53995(16)} & \num{-54080(200)}\#\tnote{a}  & \num{85(200)}\# \\
$^{106}$Zr & $0^+$  & 179(6) ms & 50, 54 & \num{0.797 085 42(4)} & \num{-58582.7(43)} & \num{-58750(200)}\#\tnote{b} & \num{167(200)}\#\\
$^{104}$Nb &  $(5^-)$ &  0.98(5) s  & 311 & \num{0.781 930 200(14)}\tnote{c} & \num{-71812.4(18)} & \num{-71811.0(18)} & \num{-1.4(25)} \\
$^{104}$Nb$^{m}$ & $(0^-,1^-)$  & 4.9(3) s  & 200 & \num{0.390 963 076(29)}\tnote{d} & \num{-71802(7)} & \num{-71801.2(19)} & \num{-0.8(73)}  \\
$^{109}$Nb & $3/2^-$  & 106.9(49) ms  & 50$^t$ & \num{0.819 672 3(18)} & \num{-56810(180)} & \num{-56690(430)} & \num{-120(470)}\\
$^{111}$Mo$^x$ & $1/2^+\#$ &  193.6(44) ms & 100 & \num{0.834 695 34(7)} & \num{-59939(9)} & \num{-59940(13)}\tnote{e} & \num{1(15)}\\
$^{112}$Mo & $0^+$ & 125(5) ms  &  50$^t$ & \num{0.842 239 6(10)} & \num{-57449(118)} & \num{-57480(200)}\#\tnote{f} & \num{31(230)}\# \\
\hline
\end{tabular}
\begin{tablenotes}
\item[a]{Also $\mathrm{ME}_{\rm lit.}$ = \num{-55213(1251)}~keV in Ref. \cite{Wang2024}.}
\item[b]{Also $\mathrm{ME}_{\rm lit.}$ = \num{-58550(173)}~keV in Ref. \cite{Knobel2016} and $\mathrm{ME}_{\rm lit.}$ = \num{-58110(696)}~keV in Ref. \cite{Wang2024}.}
\item[c]{The used accumulation time is not sufficient to resolve the ground state from the isomer, but the production is dominated by the high-spin ground state.}
\item[d]{Measured as $2^+$ ions from the in-trap-decay of $^{104}$Zr populating the low-spin isomer in $^{104}$Nb.}
\item[e]{Also $\mathrm{ME}_{\rm lit.}$ = \num{-59940(11)}~keV in Ref. \cite{Hou2023}.}
\item[f]{Also $\mathrm{ME}_{\rm lit.}$ = \num{-57470(8)}~keV in Ref. \cite{Hou2023} and $\mathrm{ME}_{\rm lit.}$ = \num{-56554(796)}~keV in Ref. \cite{Wang2024}.}
\end{tablenotes}
\end{threeparttable}
\end{table*}

The determined frequency ratios and mass-excess values (${\mathrm{ME} = (M-Au)c^2}$) are summarized in Table~\ref{tab:results}. All mass values are in agreement with AME20 \cite{AME20}. Our results constitute the first high-precision mass values of $^{104}$Y and $^{106}$Zr. The mass of $^{112}$Mo, ${\mathrm{ME} = -57449(118)}$~keV, is in agreement with the recent MR-TOF-MS measurement, ${\mathrm{ME}_{\rm lit.} = -57470(8)}$~keV, reported in Ref.~\cite{Hou2023}. The mass of $^{106}$Zr was already reported in Ref.~\cite{Knobel2016} (${\mathrm{ME}_{\rm lit.} = -58550(173)}$~keV) but rejected in the AME20 evaluation due to a significant deviation from the mass surface trends \cite{Huang2021}. Our new value of $-58582.7(43)$~keV, which differs by $167(200)$\#~keV from the AME20 extrapolation and $33(173)$~keV from Ref.~\cite{Knobel2016}, indicates that the mass surface does not follow a smooth trend.

We note that two studied nuclides, $^{104}$Nb and $^{111}$Mo, have long-lived isomeric states. A decay-spectroscopy study employing the same reaction at the IGISOL facility proposed a low-lying isomer at ${E_{x,{\rm lit.}}=100(50)}$\#~keV \cite{NUBASE20} and a half-life of about 200~ms in $^{111}$Mo \cite{Kurpeta2011}. However, only a single state was observed in this work. Considering that the phase accumulation time was 100~ms, we estimate the excitation energy to be below 60~keV. In the case of $^{104}$Nb ions produced directly in fission, the 311~ms accumulation time was not sufficient to resolve the 10-keV isomer \cite{NUBASE20} from the ground state. However, our result, ${\mathrm{ME} = -71812.4(18)}$~keV, agrees with ${\mathrm{ME}_{\rm lit.}(^{104}\mathrm{Nb}^{gs}) = -71811.0(18)}$~keV measured at the Canadian Penning Trap (CPT) \cite{Orford2018}. A second measurement was performed using $^{104}$Nb$^{2+}$ ions produced via the in-trap-decay of $^{104}$Zr, known to populate only the low-spin state in $^{104}$Nb \cite{Rinta-Antila2007}. The measured mass-excess value, ${\mathrm{ME} = -71802(7)}$~keV, is consistent with the CPT result for the isomer, ${\mathrm{ME}_{\rm lit.}(^{104}\mathrm{Nb}^{m}) = -71801.2(19)}$~keV \cite{Orford2018}. Considering that the production of high-spin states is favoured in fission \cite{Rakopoulos2019,Gao2023}, as well as the CPT results \cite{Orford2018}, we can unambiguously assign the high-spin state as the ground state and the low-spin state as the isomer in $^{104}$Nb.

\begin{figure}
\centering
\includegraphics[width=\columnwidth]{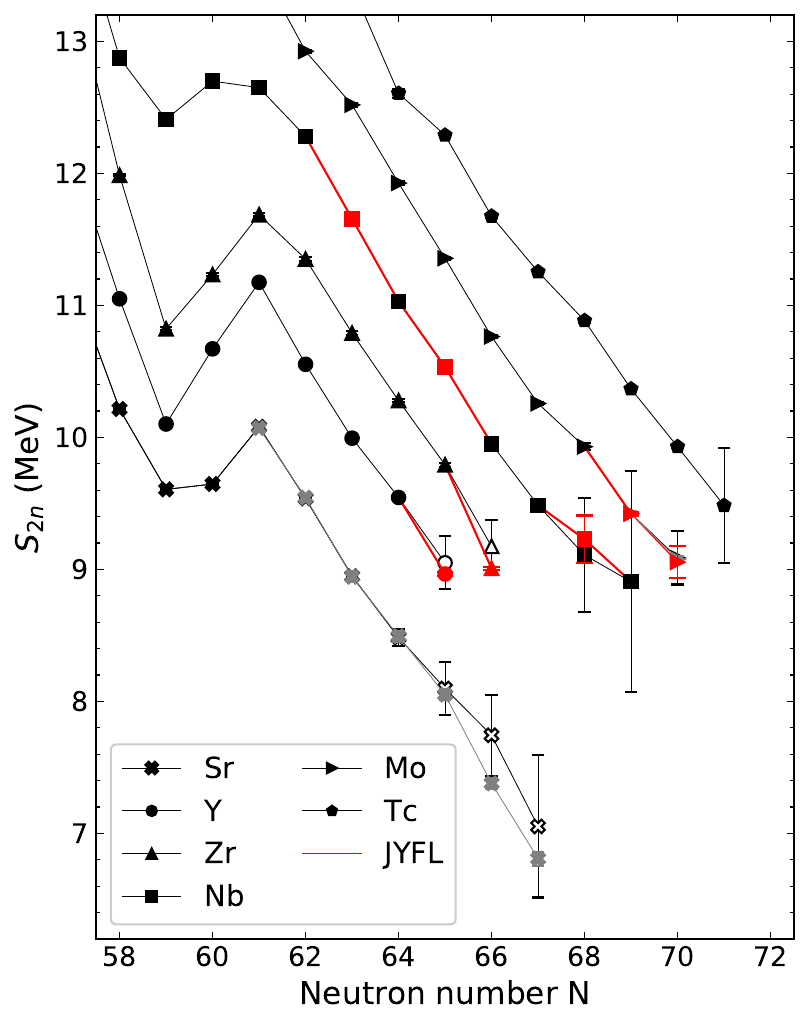}
\caption{\label{fig:S2N_JYFLTRAP+AME2020} Two-neutron separation energies $S_{2n}$ as a function of the neutron number for isotopic chains from Sr ($Z = 38$) to Tc ($Z = 43$). The values impacted by the results from this work are highlighted in red. The AME20 values are shown in black. Open symbols denote an extrapolated value in AME20. The $S_{2n}$ values with grey symbols for Sr ($Z = 38$) are from Ref.~\cite{Mukul2021} and for Mo ($Z = 42$) from Ref.~\cite{Hou2023}.}
\end{figure}

\section{\label{sec:discussion}Discussion}

The accurate reproduction of all known nuclear binding energies remains a challenge for nuclear theory. Here, 2e take this opportunity to benchmark the performance of a recent large-scale model based on an energy density functional (EDF) of the Skyrme type by comparing the masses of nuclei in this region to the ones obtained from the BSkG2 model~\cite{Ryssens2022,Ryssens2023}. Our new measurements, combined with the values of AME20, amount to 75 known experimental values for the masses for Sr, Y, Zr, Nb, Mo and Tc isotopes with ${N \geq 56}$. The local average and root-mean-square (rms) deviation of the masses of these nuclei amount to ${\bar{\epsilon}_{\rm local} = -0.447}$~MeV\footnote{We define the deviation as (theory - experiment), such that the sign of $\bar{\epsilon}$ indicates the model overbinds these nuclei on average, i.e. produces mass excesses that are too large in absolute size.} and ${\sigma_{\rm local} =0.866}$~MeV, respectively. The rms deviation in this region is somewhat larger than the global accuracy of the model (${\sigma_{\rm global} = 0.687}$~MeV), largely because of the non-zero (local) value of $\bar{\epsilon}$. These values do not necessarily point to a specific modelling deficiency with respect to this region of the nuclear chart, but rather reflect the typical currently unavoidable compromise achieved by fitting the model parameters to thousands of known nuclear masses. Nevertheless, other models used to study the region do not even report on their accuracy for absolute masses~\cite{deRoubin2017}.

Compared to absolute values, mass differences often indicate more clearly whether a model manages to capture the structural evolution along isotopic or isotonic chains. Two-neutron separation energies $S_{2n}$ are typically used as an indicator for structural changes. Figure~\ref{fig:S2N_JYFLTRAP+AME2020} shows $S_{2n}$ values for isotopic chains ranging from Sr ($Z=38$) to Tc ($Z=43$). While our new experimental data agree with the AME20 values \cite{AME20}, the improved precision obtained in this work clarifies the trends in the $S_{2n}$ values in the studied isotopic chains. Two interesting features emerge when crossing the midshell at $N=66$. First, the new values for $^{104}$Y ($N=65$) and $^{106}$Zr ($N=66$) fall below a straight linear extrapolation of AME20 in both the Y ($Z=39$) and Zr ($Z=40$) isotopic chains, similarly to the Sr chain recently studied at TITAN~\cite{Mukul2021}. Second, the new value for $^{109}$Nb ($N=68$) is above the linear trend. As a result, the gap between the isotopic chains below and above $Z=40$ seems to increase when approaching the neutron midshell region at $N\approx66$.

\begin{figure*}
\centering
\includegraphics[width=\textwidth]{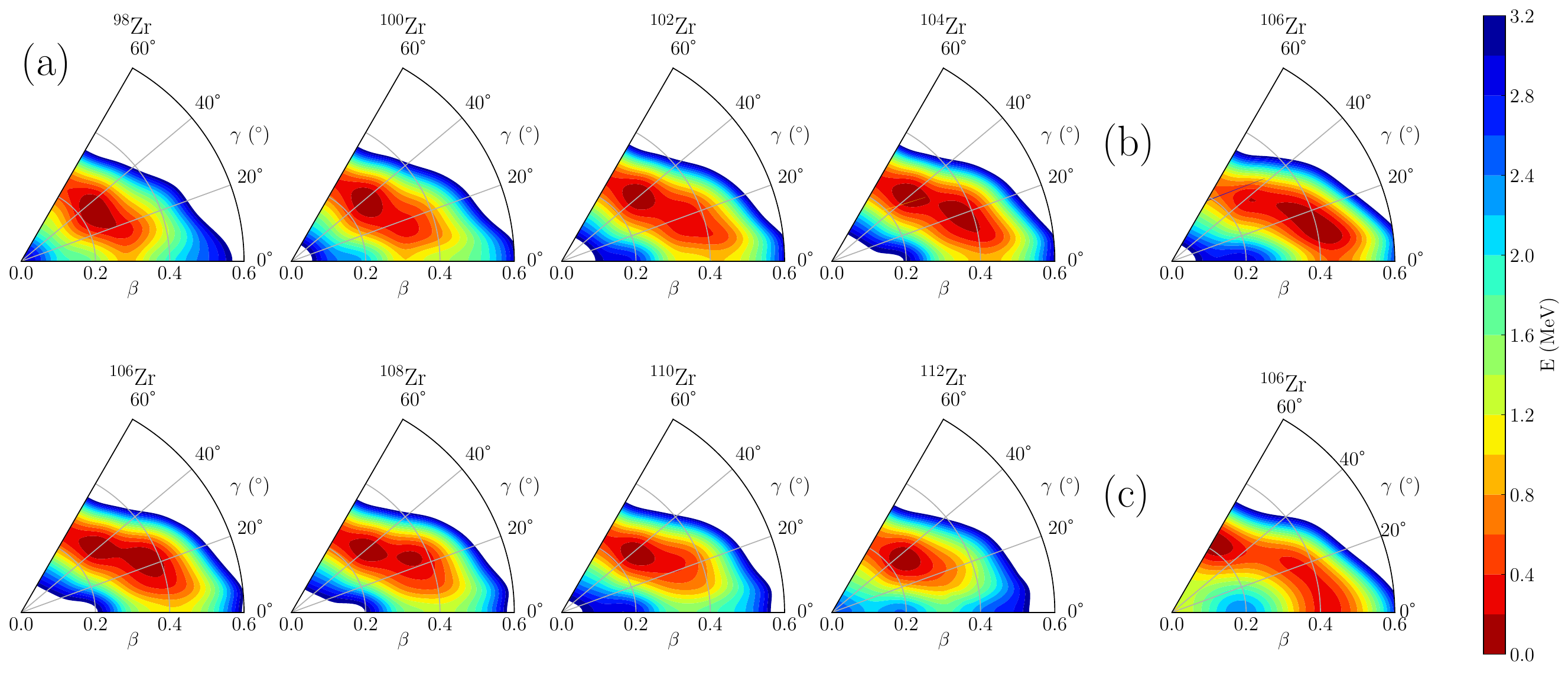}
\caption{\label{fig:106ZrPES} 
Potential energy surfaces (PESs) of even-even Zr isotopes in the $(\beta,\gamma)$ plane as obtained with BSkG2, all normalized to the respective minimum. Panel (a): total energy for $^{98-108}$Zr. Panel (b): total energy minus the vibrational correction for $^{106}$Zr. Panel (c): total energy minus the rotational and vibrational correction for $^{106}$Zr. See text for details.}
\end{figure*}

In Fig.~\ref{fig:106ZrPES}(a) we show the potential energy surfaces (PESs) for even-even neutron-rich Zr isotopes as a function of $\beta$ and $\gamma$ (see Ref.~\cite{Scamps2021} for precise definitions).
For even-even $^{98-108}$Zr isotopes, there is always a triaxial minimum (T1) at modest total deformation $\beta_2 \approx 0.2-0.25$ and at high values of $\gamma \approx 40^{\circ}$. For $^{104,106,108}$Zr, there is a second triaxial minimum (T2) very close in energy ($\approx$ 150 keV energy diﬀerence) that is separated by a very thin barrier from T1. These second triaxial minima have a larger deformation of $\beta_2 \approx 0.35$ and smaller $\gamma \approx 20^{\circ}$. 

For even-even nuclei, this type of PESs with two competing minima that are both triaxial is rarely found in mean-field calculations. This structure appears here because of the inclusion of so-called rotational and vibrational corrections in BSkG2, which aim at eliminating the spurious energy associated with the corresponding type of collective motion that arises from a symmetry-broken mean-field description of the nucleus. For BSkG2, these corrections take the form of simple analytical expressions of the (rotational) moment of inertia of the nucleus and include 25 parameters in the model optimisation protocol. Both recipes are essentially phenomenological albeit to differing degrees: the rotational correction has a microscopic foundation in many-body theory and is widely used \cite{Tondeur2000,Bender2004,Goriely2013}, while the ansatz for the vibrational correction is more ad-hoc~\cite{Ryssens2023}. Both corrections are strongly deformation dependent, meaning that their inclusion can significantly impact the PES. We illustrate this for $^{106}$Zr in the other panels of Figure~\ref{fig:106ZrPES}. Without the vibrational correction (see Fig.~\ref{fig:106ZrPES}(b)), the T2 minimum becomes the lowest in energy and moves to a somewhat larger deformation. If both collective corrections are removed (see Fig.~\ref{fig:106ZrPES}(c)), the PES becomes more traditional: the triaxial minima get replaced by coexisting axially symmetric minima, one oblate and one prolate.\footnote{Interestingly, D1S predicts PES for $^{106}$Zr and $^{108}$Zr that have three coexisting minima: one prolate, one oblate and a very shallow triaxial one with a deformation in between that of the others~\cite{hilaire_amedee_2007,Delaroche2010}. These calculations also feature a zero-point-energy correction that is similar in spirit to our collective corrections.} 
This behavior is in qualitative agreement with the available EDF results that use more advanced techniques to include these effects \cite{Schunck2019} which also tend to transform axial minima into triaxial ones \cite{Paul2017,Rodriguez11}.

\begin{figure}[h!t!b]
\centering
\includegraphics[width=\columnwidth]{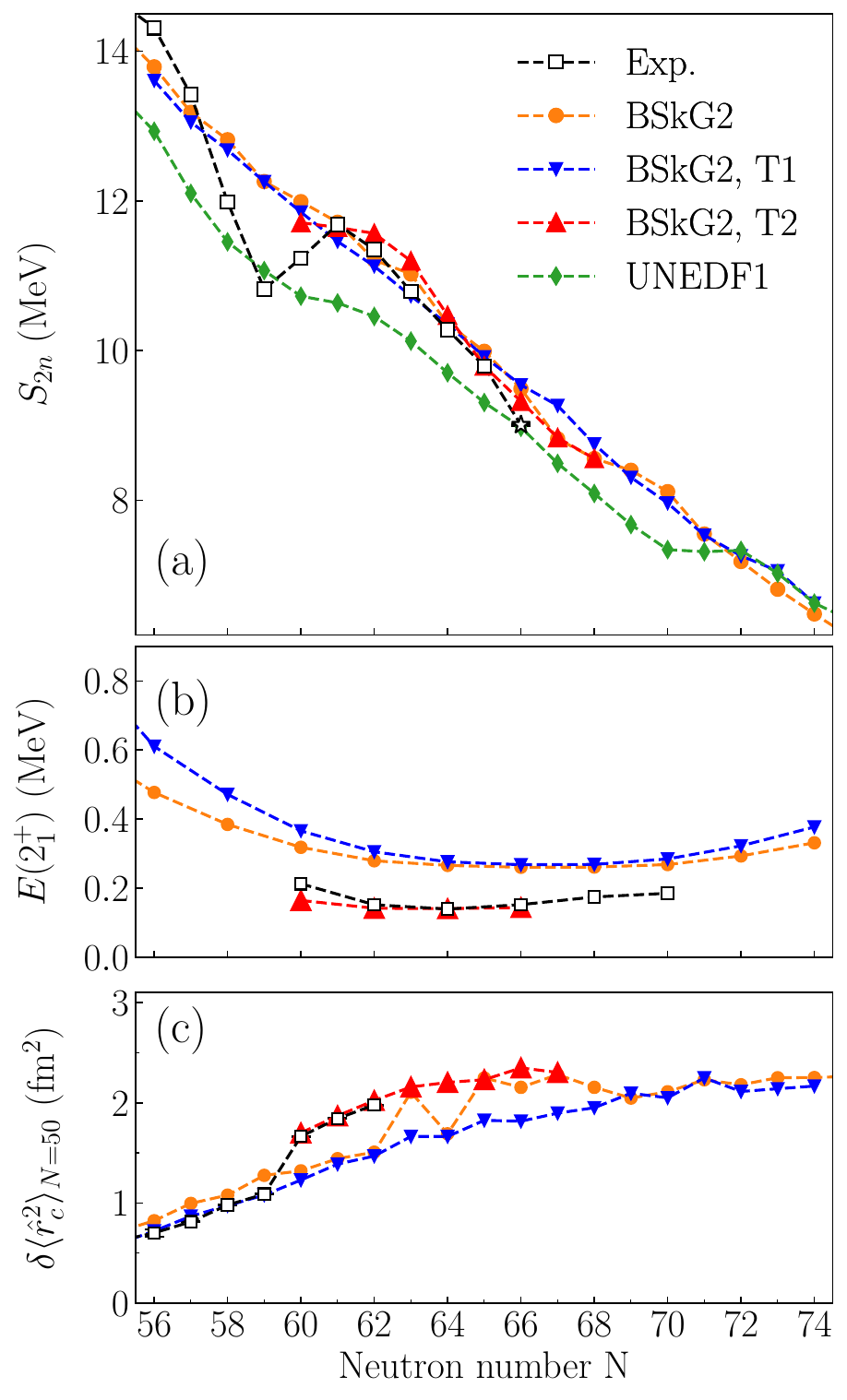}
\caption{\label{fig:exptheory} Experimental data for the Zr isotopes compared to the BSkG2 model, as well as to the BSkG2 results obtained without vibrational corrections for the T1 and T2 triaxial minima (see text for details). Panel (a): Two-neutron separation energies. For comparison, the $S_{2n}$ values from the UNEDF1 \cite{Kortelainen2012} EDF model taken from Ref. \cite{Erler12} are also shown. Panel (b): Excitation energies of the first $2^+$ state. Panel (c): The mean-square charge radii $\delta\langle r^2\rangle$ with respect to $^{90}$Zr. The experimental $E(2^+_1)$ values for the $N < 60$ Zr isotopes are not presented as they lie above 1 MeV and are non-rotational. Experimental values for the $S_{2n}$ are from this work for $^{106}$Zr (highlighted with a star marker) and AME20 \cite{AME20} while the values of $E(2^+_1)$ and $\delta \langle \hat{r}_c^2 \rangle$ were taken from Refs.~\cite{ENSDF,Paul2017} and Ref.~\cite{angeli2013}, respectively.}
\end{figure}

We compare the calculated two-neutron separation energies to the experimental results along the Zr isotopic chain in Fig.~\ref{fig:exptheory}(a). The trend of the BSkG2 model is very linear near $N=60$ and $N=66$. It misses the new experimental point at $^{106}$Zr and the pronounced ``dip'' in the experimental values around at $N{\approx59}$. The latter is not reproduced - to the best of our knowledge - by any EDF-based mean-field model, as we illustrate with the Skyrme-type UNEDF1 parameterization and as can be seen for the D1S and D1M Gogny parameterizations in Ref.~\cite{chimanski2023}. Despite these deficiencies, BSkG2 captures the general trend of the $S_{2n}$ values, especially when compared to the UNEDF1 parameterization~\cite{Kortelainen2012} which was fitted to less masses.

The vibrational correction is arguably the most phenomenological ingredient in our calculations, yet it biases the PES in favor of the minimum with smallest deformation in the Zr isotopic chain. To investigate how the selection of minima impacts the calculated observables we remove the vibrational correction and plot in Fig.~\ref{fig:exptheory}(a) the $S_{2n}$ values for the resulting T1 and T2 minima. For the two-neutron separation energies, the general trend is rather similar to the unmodified BSkG2 calculation. However, some differences emerge. First, the T2 minimum without vibrational correction is closer to the new experimental value for $^{106}$Zr. Second, the trend of the T1 minima is extremely linear while the T2 curve shows some structure near $N=59$ that vaguely resembles the experimental dip in shape, even if clearly not in amplitude. The reproduction of the experimental feature at $N=59$ is hardly satisfactory but other attempts to fine-tune BSkG2 were even less successful. We take this as an indication that a mean-field description is not rich enough to simultaneously describe the dip of the $S_{2n}$ and other observables in this mass region, especially given the similar deficiencies of virtually all earlier mean-field results.

Although the improvement in the description of the $S_{2n}$ values is minor, removing the vibrational correction and selecting the T2 minima does lead to a much improved description of two other properties of the Zr isotopes for which a T2 minimum is found with BSkG2. The excitation energies of the first $2^+$ states are compared to the available data in Fig.~\ref{fig:exptheory}(b). The model values are those of a perfect rigid triaxial rotor characterized by the BSkG2 moments of inertia, calculated microscopically and renormalized by the traditional factor 1.32~\cite{libert1999}. 
Note that the first $2^+$ states of Zr isotopes with $N < 60$ lie above 1 MeV and are certainly non-rotational. This is also supported by the strong increase in the ratio of the $4_1^+/2_1^+$ energies between $^{98}$Zr and $^{100}$Zr~\cite{ENSDF}. Thus, experimental values for these isotopes are not plotted. Like the original full calculation, results for the T1 minimum without vibrational correction that is found for all isotopes overestimate the excitation energies. By contrast, the results for the T2 minimum without vibrational correction reproduce this quantity between $N=60$ and $N=66$, which coincides with the full range of even-even isotopes for that this minimum is found.
The second property is the mean-square charge radius $\delta \langle \hat{r}_c^2 \rangle$ relative to $N=50$, see Fig.~\ref{fig:exptheory}(c). Without the vibrational correction, the T2 minimum agrees well with the available data~\cite{Campbell2002,angeli2013} that are limited to $N \leq 62$. In particular, the sudden increase at $N=60$ is well reproduced, a feature that in general is difficult to find at the precise neutron number~\cite{manea_penning-trap_2017,sels_shape_2019}. 
Unmodified BSkG2 calculations fail to match this transition and furthermore produce slightly too large shifts for $56 \leq N \leq 59$. For more neutron-rich isotopes, the competition between both minima results in a staggering at $N \approx 64$. Beyond
$N=67$, BSkG2 does not predict a pronounced T2 minimum anymore, in contradiction with the low $E(2^+_1)$ energy of $^{108}$Zr and $^{110}$Zr. It has already been conjectured in Ref.~\cite{Paul2017} that the available spectroscopic information for $^{110}$Zr points towards a larger deformation than what is predicted by the mean-field models used in that study.

In axial calculations, the UNEDF1 parameter set has similar problems to reproduce the deformation effects in neutron-rich Zr isotopes~\cite{Erler12}: ground state deformation is only found for even-even isotopes between $N=60$ and $N=68$, with $N=60$, 66 and 68 having small oblate deformation and $N=62$ and 64 having large prolate deformation. Similar results with differences in detail are also found for D1S~\cite{chimanski2023}. The new data corroborate the difficulties of present mean-field approaches to precisely describe the evolution of near-degenerate coexisting deformed minima in the Zr isotopes, similar to what is also found
in the neutron-deficient Pb region~\cite{manea_penning-trap_2017,sels_shape_2019}.

We note in passing that, together with the $E(4_1^+)/E(2_1^+)$ ratio of about 3, the low-lying $2^+_1$ level of $^{110}$Zr rules out the possibility of an exotic spherical harmonic oscillator shell closure and also of an even more exotic tetrahedrally deformed shell closure at $N=70$ that were earlier predicted in the literature for this neutron-rich Zr isotope, see Ref.~\cite{Paul2017}.

The mean-square charge radii and $2^+_1$ excitation energies are directly connected to deformation. For these quantities a simple mean-field modelling quite accurately describes the experimental data, provided one discards the phenomenological vibrational correction that biases the competition between both minima in the wrong direction. These modifications also appear as a limited step in the right direction for the two-neutron separation energies. Although other modifications of the model could theoretically be used to change the balance between the minima, our discussion illustrates that phenomenological corrections for collective motion can locally bias models in regions of the nuclear chart where multiple shapes compete. Similar problems, however, can also be found when combining a more microscopic treatment of the corrections with traditional parametrisations of the EDF \cite{Paul2017}. A study of the impact of the collective corrections on different observables in different areas of the nuclear chart would be welcome, as would be improved ways to include collective motion in large-scale models. Ideally, the ongoing development of beyond-mean-field techniques~\cite{Schunck2019} should be pushed to the point where their use at the scale of the nuclear chart becomes feasible for nuclei with even and odd numbers for protons and neutrons.

\section{\label{sec:conclusions}Conclusions}

In this Letter, we reported on precision mass measurements of neutron-rich refractory isotopes that pin down the trends of the mass surface in the $N \approx 66$ midshell region. The two-neutron separation energies in both the Y and Zr isotopic chains decrease more steeply than a linear extrapolation of previously known values toward $N=66$. The recent BSkG2 model reproduces the absolute masses in the region within about 900 keV, compatible with a typical accuracy of today's large-scale models. However, it entirely misses the interesting feature of the two-neutron separation energies near $N=60$. Focusing on the Zr isotopic chain, BSkG2 predicts a competition between two close-lying minima that are both triaxial. Which minimum is preferred by a large-scale model that is restricted to the mean-field level depends sensitively on several factors. We have shown in particular that it depends on the correction for spurious collective motion.  Manually selecting the more deformed of the minima in calculations and removing the vibrational correction, the calculations are more in line with experimental data on mean-square charge radii and $E(2_1^+)$ energies and result in a non-linear trend of the two-neutron separation energies trend near $N=60$ that carries some resemblance to experimental feature.

We conclude that the simple recipes used to account for spurious motion in large-scale models are too crude to accurately capture all details of the mass surface in regions of the nuclear chart where multiple configurations compete. Nevertheless, such recipes are crucial for the globally accurate reproduction of masses and other observables. This work highlights that such global success comes at the cost of local failure and that there is a need for an improved treatment of collective motion in large-scale models, ideally through beyond-mean-field techniques.

\section*{Acknowledgments}

The present research benefited from computational resources made available on the Tier-1 supercomputer of the F\'ed\'eration Wallonie-Bruxelles, infrastructure funded by the Walloon Region under the grant agreement No. 1117545. This project has received funding from the European Union’s Horizon 2020 research and innovation programme under grant agreements No. 771036 (ERC CoG MAIDEN) and No. 861198–LISA–H2020-MSCA-ITN-2019 and the Academy of Finland projects No. 295207, 306980, 327629, 354968 and 354589. W.R. is a Research associate of the F.R.S.-FNRS (Belgium). Work by M.B. has been supported by the Agence Nationale de la Recherche, France, Grant No.~19-CE31-0015-01 (NEWFUN). We are grateful for the mobility support from Projet International de Coop\'eration Scientifique Manipulation of Ions in Traps and Ion sourCes for Atomic and Nuclear Spectroscopy (MITICANS) of CNRS. 

\bibliographystyle{elsarticle-num} 
\bibliography{biblio}

\end{document}